\documentclass[final,3p,times,10pt]{elsarticle}

\usepackage{multirow,setspace,times,amssymb,amsmath,graphicx,color,rotating,subfigure,url}
\usepackage{lineno,color}
\usepackage{natbib}
\usepackage{booktabs}%
\usepackage{longtable}%
\graphicspath{{Figures/}}
\usepackage[table]{xcolor}
\usepackage{tabularx}
\usepackage{graphicx} 
\usepackage{epstopdf}
\usepackage{mathrsfs}
\usepackage{makecell}
\usepackage[bookmarks=true,colorlinks,linkcolor=blue,anchorcolor=blue,citecolor=blue,unicode]{hyperref}
\usepackage{bookmark}
\usepackage{setspace}

\hypersetup{CJKbookmarks=true}%
\journal{}
\begin{document}

\begin{frontmatter}
\title{An interpretable machine-learned model for international oil trade network}

\author[SB,RCE]{Wen-Jie Xie}
\author[SB]{Na Wei}
\author[SB,RCE,DM]{Wei-Xing Zhou\corref{WXZ}}
\ead{wxzhou@ecust.edu.cn}
\cortext[WXZ]{Corresponding author. Corresponding to: 130 Meilong Road, P.O. Box 114, School of Business, East China University of Science and Technology, Shanghai 200237, China.}

\address[SB]{School of Business, East China University of Science and Technology, Shanghai 200237, China}
\address[RCE]{Research Center for Econophysics, East China University of Science and Technology, Shanghai 200237, China}
\address[DM]{Department of Mathematics, East China University of Science and Technology, Shanghai 200237, China}


\begin{abstract}
Energy security and energy trade are the cornerstones of global economic and social development. The structural robustness of the international oil trade network (iOTN) plays an important role in the global economy. We integrate the machine learning optimization algorithm, game theory, and utility theory for learning an oil trade decision-making model which contains the benefit endowment and cost endowment of economies in international oil trades. We have reconstructed the network degree, clustering coefficient, and closeness of the iOTN well to verify the effectiveness of the model. In the end, policy simulations based on game theory and agent-based model are carried out in a more realistic environment. We find that the export-oriented economies are more vulnerable to be affected than import-oriented economies after receiving external shocks. Moreover, the impact of the increase and decrease of trade friction costs on the international oil trade is asymmetrical and there are significant differences between international organizations.
\end{abstract}


\end{frontmatter}


\section{Introduction}
\label{S0:Introduction}

Energy security and energy cooperation are vital factors affecting global economic stability and development. In the face of natural disasters, local wars, and climate changes, how to ensure energy security and stable energy cooperation is an issue that needs to be addressed \citep{Li-Dong-Jiang-Dong-Liu-2021-ResourPolicy,Xi-Zhou-Gao-Liu-Zheng-Sun-2019-EnergyEcon,Caraiani-2019-EE,Sun-An-Gao-Guo-Wang-Liu-Wen-2019-Energy}. The distribution of petroleum resources is markedly imbalanced. The large-scale and long-distance transportation has caused tight capacity and increased costs, which have affected the coordinated development of the energy industry \citep{Li-Gao-An-Zheng-Wu-2021-ResourPolicy,Zhang-Ji-Fan-2015-EnergyEcon,Le-Chang-2013-EE,Rafiq-Sgro-Apergis-2016-EE}. In addition to the fragile global energy supply-demand balance and frequent oil market fluctuations, local wars and sanctions, international oil price shocks and various other non-economic factors also affect international cooperation in energy \citep{An-Wang-Qu-Zhang-2018-Energy,Du-Wang-Dong-Tian-Liu-Wang-Fang-2017-AEn,Du-Dong-Wang-Zhao-Zhang-Vilela-Stanley-2019-Energy}. How to solve these problems requires cross-disciplinary cooperation. The remarkable development of artificial intelligence technologies and methods have provided new tools and methods for addressing the problem of irrational energy structure and for expanding international energy cooperation.

In recent years, the popularity of artificial intelligence and machine learning algorithms has extended to economics and finance research \citep{Yu-Wang-Lai-2008-EE,Franke-Hoser-Schroder-2008-bk,Silver-Hubert-Schrittwieser-Antonoglou-Lai-Guez-Lanctot-Sifre-Kumaran-Graepel-Lillicrap-Simonyan-Hassabis-2018-Science}. Increasingly, scholars adopt machine learning algorithms to analyze and solve the complex problems encountered in their respective research areas, achieving many good results and encountering 
considerable challenges \citep{Varian-2014-JEP}. For example, deep learning in economic forecasting suffer from a well-known and criticized black box problem. Even though deep learning algorithms can improve the accuracy of predictions, the economic implications of the model cannot be reasonably explained \citep{Athey-2018-NBER}. The scholars in the field of financial economics are very much focused on analyzing the causal relationships between variables, rather than just being satisfied with the correlations found by most deep learning algorithms \citep{Jasny-Stone-2017-Science}. However, many scholars 
believe that causal relationships and correlations can be combined for analysis \citep{Athey-Imbens-2019-ARE,Yuan-Alabdulkareem-Pentland-2018-NC}. While machine learning methods are mostly used to accomplish prediction tasks in the field of economics \citep{Zhong-Enke-2019-FinancInnov,Lahmiri-2014-FNL,Wang-Huang-Wang-2013-NCA}, the biggest highlight of this paper's approach is to provide interpretable economic models, which have very large application scenarios, by fitting interpretable model parameters through machine learning optimization methods.

Although the application of machine learning algorithms in the economic field has certain limitations, scholars have achieved many research results in recent years by using machine learning methods to study problems in the economic field \citep{Ghoddusi-Creamer-Rafizadeh-2019-EE,Voyant-Notton-Kalogirou-Nivet-Paoli-Motte-Fouilloy-2017-RE,Zeng-Liu-Su-Hu-2018-ANE}. Zhao et al. use machine learning methods to predict energy prices and evolutionary trends \citep{Zhao-Li-Yu-2017-EE,Kleinberg-Ludwig-Mullainathan-Obermeyer-2015-AER}. In addition, the study of energy consumption or energy supply and demand based on machine learning methods is also a hot research topic \citep{Tang-Yu-Wang-Li-Wang-2012-AE}. How 
to produce convincing results applying machine learning methods has always been a concern in machine learning and computer science. As a result, a large number of studies and tools addressing the interpretability of deep learning algorithms have also emerged \citep{Yuan-Alabdulkareem-Pentland-2018-NC}. We believe that more practical machine learning tools will be widely used in the economic field \citep{Athey-2017-Science}.

In a complex trading environment, economies' decision-making is affected by many factors. How to represent these heterogeneous economies' attributes and environmental factors is an important issue. In different decision-making environments, the factors that affect the attributes of the economies are different, and many factors are unquantifiable due to privacy and difficulty of collection. We introduce a representation method of heterogeneous economies in complex environments \citep{Yuan-Alabdulkareem-Pentland-2018-NC}. Based on the decision-making process of economies, we integrate machine learning optimization algorithms with the oil trade network to directly learn and reproduce the decision-making process of the economies. In order to simulate the oil trade system more realistically and provide more practical strategic support for governments and organizations in decision-making, we provide a new analysis framework to directly learn the representation and decision-making process of the economies from the iOTNs, and then integrate game theory with the agent-based model, and finally achieve the purpose of policy simulation and evaluation.

The main contributions can be summarized in three respects: 1) We construct a heterogeneous economies decision-making model based on utility theory and game theory, considering the endowments of heterogeneous economies in oil trade.
2) We use machine learning optimization algorithms to learn the endowment feature vectors of heterogeneous economies and examine the interpretability of model parameters.
3) Under the ``scenario" of trade friction changes, we explore the evolution of economic entities. For example, policy simulations and policy evaluations are conducted, considering the impact of trade war and the COVID-19 epidemic.

This article is organized as follows. Section \ref{S1:LitRev} is a literature review. Section \ref{S1:Data:Methodology} introduces the research data and methods. Section \ref{S1:EmpAnal} uses international oil trade data to construct iOTN, and conduct empirical analysis based on the relevant methods in section \ref{S1:Data:Methodology}. Section \ref{S1:PolicySimulation} conducts a policy simulation on the iOTN. Section \ref{S1:Conclude} is discussion and application.

\section{Literature review}
\label{S1:LitRev}

Energy is the blood of an economic system, and oil is an essential energy resource. Since the oil resources distribution is uneven, economies need trading to balance oil supply and demand \citep{Xi-Zhou-Gao-Liu-Zheng-Sun-2019-EnergyEcon,Caraiani-2019-EE,Sun-An-Gao-Guo-Wang-Liu-Wen-2019-Energy}. As a result, economies interact and become inextricably intertwined because of the complex oil trade relationship \citep{Liu-Chen-Wan-2013-EM}. The oil trading system, which is crucial to the development of the global economy, has received considerable attention from scholars \citep{Fagiolo-Reyes-Schiavo-2009-PhysRevE,Fagiolo-2010-JEconInteractCoord,Dablander-Hinne-2019-SR}. As a complex system, the oil trading system  finds complex network methods suitable for and analysis \citep{Battiston-Farmer-Flache-Garlaschelli-Haldane-Heesterbeek-Hommes-Jaeger-May-Scheffer-2016-Science,Haldane-May-2011-Nature}. We can integrate international oil trade into a complex system. In the modeling process, each energy economy can be regarded as a node, and the trade relationship between economies can be abstracted as a network connection \citep{Bhattacharya-Mukherjee-Saramaki-Kaski-Manna-2008-JStatMech,Zhang-Lan-Xing-2018-IOP,Hao-An-Sun-Zhong-2018-ResourPolicy,Li-Dong-2020-ResourPolicy}.

The formation of the iOTN depends on the development of the economies and their trade relations. The international oil trading system is evolving into a stable, orderly, and integrated system \citep{An-Zhong-Chen-Li-Gao-2014-Energy,Yu-Jessie-Sharmistha-2015-Energy}. Many factors affect the formation of the oil trade network. Zhang et al. used spatial econometric models to examine significant factors such as supply and demand, technological progress, and energy efficiency \cite{Zhang-Ji-Fan-2015-EnergyEcon}. Zhang et al. investigated the influence of competition and dependence between economies and geographic factors on the oil trade \cite{Zhang-Ji-Fan-2014-EP,Kharrazi-Fath-2016-EP}. Kitamura et al. studied the gravity equation and described the flow of oil trade, pointing out that the bilateral trade volume is directly proportional to the total production value of the two sides and inversely proportional to the distance between them \cite{Kitamura-Managi-2017-AEn}. Some scholars also studied each economy's status in the iOTNs \citep{Du-Wang-Dong-Tian-Liu-Wang-Fang-2017-AEn,Zhong-An-Shen-Fang-Gao-Dong-2017-EP}, and analyzed the influence of different factors on the oil trade of economies. The oil trade relationships are extraordinarily complex and are affected by many observable and unobservable factors.

The oil trade relationships are affected not only by the attributes of the economies but also by the structure of the oil trade network. Gomez et al. used the traditional gravity model when studying trade network relations \citep{Gomez-Herrera-2013-EmpE}. The gravity model is also used to discover the potential trade flows of agricultural products \citep{Shuai-2010-OutlookAgric,Ravishankar-Stack-2014-WE}. Feng et al. applied the prediction method of potential trade relations to the natural gas market \citep{Feng-Li-Qi-Guan-Wen-2017-Energy,Guan-An-Gao-Huang-Li-2016-Energy}. In most of the literature, the factors affecting trade relations cannot be fully examined, and it is almost an impossible task. To give more considerations to multiple factors such as individual attributes and trade network structure, we have adopted a different approach. First, it is assumed that the existing trade network is influenced by almost all the factors whose information is also embedded in the oil trade network. We introduce a machine learning method similar to reverse engineering, learning optimization algorithms and the endowment representation of the economies in the oil trade network through the machine.  Then, we can reproduce the oil trade decision-making process of the economy and predict the trade network relationship. Finally, the network evolution can be simulated and analyzed, and the network evolution index can be quantified, such as network effectiveness \citep{Latora-Marchiori-2001-PRL,Sheng-Wu-Shi-Zhang-2015-EnergyEcon,Xie-Wei-Zhou-2020-JStatMech,Garlaschelli-DiMatteo-Aste-Caldarelli-Loffredo-2007-EurPhysJB}. 

\section{Data and methodology}
\label{S1:Data:Methodology}

\subsection{Construction of iOTN}

The global oil trade data from 1990 to 2019 comes from UN Comtrade, and the oil data code is HS270900. The data comes from the official data reported by both sides of the trade.The United Nations Comtrade database aggregates detailed global annual and monthly trade statistics by product and trading partner for use by governments, academia, research institutes, and enterprises. Data compiled by the United Nations Statistics Division covers approximately 200 countries and represents more than 99\% of the world's merchandise trade. There are many studies~\citep{An-Zhong-Chen-Li-Gao-2014-Energy,Kharrazi-Fath-2016-EP,Zhong-An-Shen-Fang-Gao-Dong-2017-EP,An-Wang-Qu-Zhang-2018-Energy,Du-Dong-Wang-Zhao-Zhang-Vilela-Stanley-2019-Energy} have used the data to explore the international oil trade, with instructive results. The oil trading economies refer to almost all the countries or regions around the world involved in oil trading. This study uses more complete oil trade import data \citep{Fan-Ren-Cai-Cui-2014-EconModel}. We extract the required data from the original data, including the data of time, exporting economies, importing economies, and transaction volumes (converted into US dollars according to the UN Statistics standard). 

\begin{figure}[!t]
\centering
\includegraphics[width=0.99\linewidth]{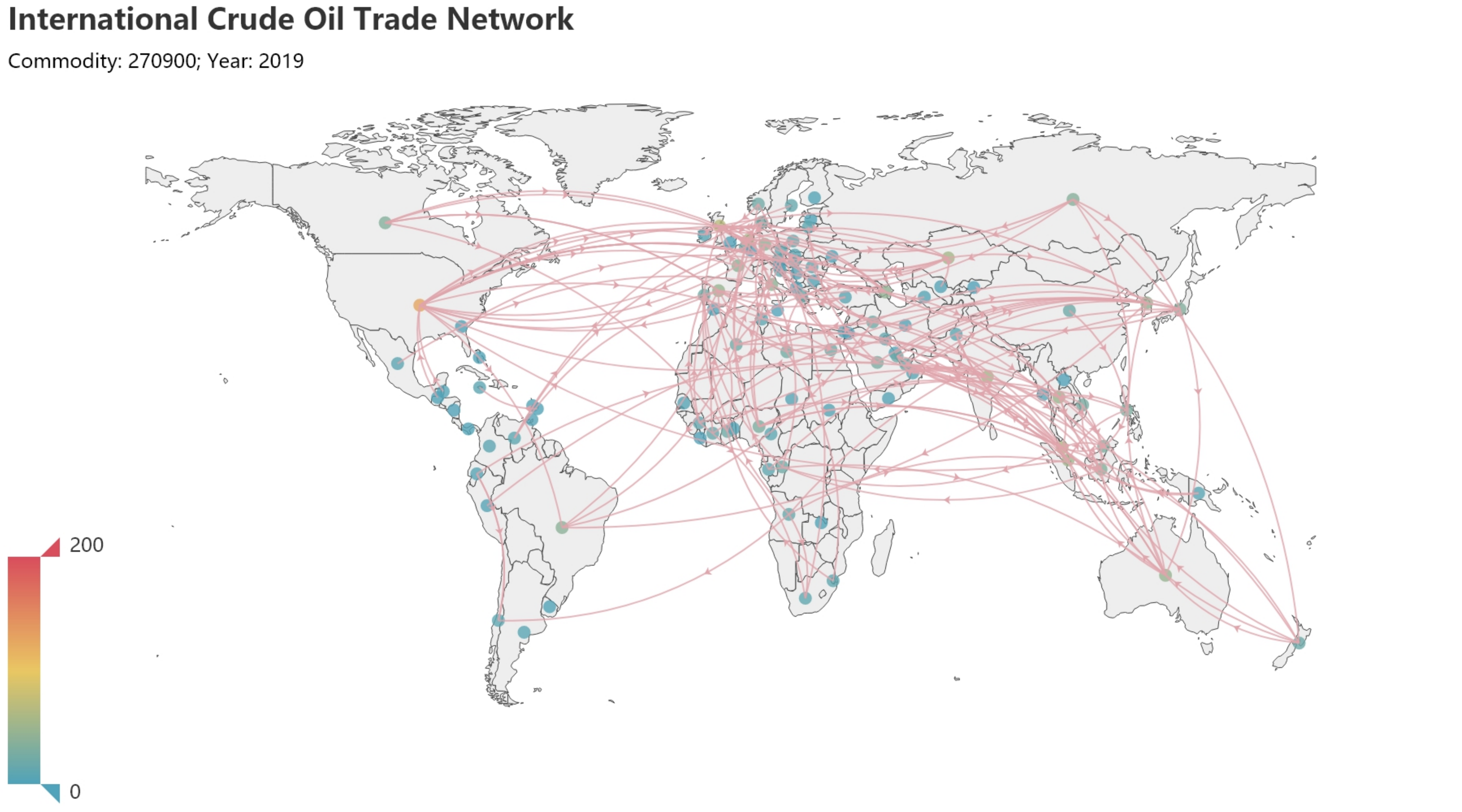}\\
  \caption{
  An example of the topology of an iOTN. For clarity, we randomly display 20\% of trade relations.
  }
    \label{Fig:oil:cyto}
\end{figure}

The global oil trade market can be abstracted as an iOTN, with economies as network nodes and trade relations as network connections. We construct weighted oil trade networks from 1990 to 2019. The iOTN can be represented by the matrix $\mathbf{W}(t)=[w_{ij}(t)]$, and $w_{ij}(t)+w_{ji}(t)$ denotes the sum of the import and export trade volume between the economy $i$ and $j$ in the year $t$. In order to remove some noise, it is necessary to filter the network edges and extract the backbone of the network. The simplest way to filter is to set a transaction volume as a threshold and delete trade relationships with fewer volumes than the threshold. We use the threshold filtering method to construct the skeleton of the iOTN and obtains the adjacency matrix of the oil trade network $\mathbf{A}=[a_{ij}]$, and the adjacency matrix element $a_{ij}=a_{ ji}=1$ means that the trade volume $w_{ij}(t)+w_{ji}(t)$ between economy $i$ and economy $j$ exceeds one million U.S. dollars in at least one of the 30 years. Figure \ref{Fig:oil:cyto} is a schematic diagram of the topology of the iOTN. We only considered the most basic trade relation filtering method and did not compare the impact of different network edge filtering methods on the results.

\subsection{Decision-making model of international oil trade}

In the trade network, the formation of oil trade relations between economies needs to consider many trade factors, such as oil prices, oil import and export routes, energy production, and economic culture. Many factors affect the stability and vulnerability of trade relations between economies. In order to characterize the oil trade relationship between economies, we use a endowment vector to represent these trade-related factors, each dimension of the vector representing an attribute. The endowment vector between the two economies is used as decision variables when establishing oil trade relations. The benefits and costs of oil trade relations can be measured by utility functions \citep{Yuan-Alabdulkareem-Pentland-2018-NC}. The utility function of the economy $i$ is defined as follows:
\begin{equation}
   U_i(S; {\mathbf{E}}, {\mathbf{b}}, {\mathbf{c}})=F_i(S; {\mathbf{E}}, {\mathbf{b}})-G_i(S; {\mathbf{E}}, {\mathbf{c}}), ~~\forall S\subset{\mathcal{I}/i},
   \label{Eq:Ui}
\end{equation}
where $S$ is the trading partners of $i$, and $\mathcal{I}$ is the set of economies. $\mathcal{I}/i$ represents the set of economies that does not include the economy $i$.
The matrix $\mathbf{E}$ is the endowment of all economies in the oil trade network. Each row of the matrix corresponds to an endowment vector of an economy, and each column corresponds to an endowment dimension, which represents a trade factor. ${\mathbf{b}}$ stands for the importance of the benefit attribute. The benefit endowment refers to factors such as resource reserves. The trade between economies with large differences in resource reserves can bring benefits. In addition, ${\mathbf{c}}$ is the importance of cost attributes. The cultural differences and language barriers will increase trade costs, and the costs are not conducive to trade relations. Therefore, the difference in the cost endowment in the trade relationship will cause a large trade loss \citep{Yuan-Alabdulkareem-Pentland-2018-NC}. The utility function Eq.~(\ref{Eq:Ui}) distinguishes between the benefit endowment and the cost attribute. The benefit endowment is the first $D_b$ column in the $\mathbf{E}$ matrix. The cost-related endowment is the last $D-D_b$ columns of the matrix. As stated by Yuan et al., we also call the matrix $\mathbf{E}$ as endowment matrix \citep{Yuan-Alabdulkareem-Pentland-2018-NC}. When the energy endowment of the economy $j$ is greater than the economy $i$, the economy $i$ will trade with the economy $j$. The matrix $\mathbf{E}$, weight coefficients ${\mathbf{b}}$ and ${\mathbf{c}}$ are learnable variables \citep{Yuan-Alabdulkareem-Pentland-2018-NC}, and the benefit function is recorded as
\begin{equation}
   F_i(S^{*};{\mathbf{E}}, {\mathbf{b}})= \sum_{i\in{S_i^{*}}}\sum_{d=1}^{D_b} b_d\max(e_{jd}-e_{id},0).
   \label{Eq:Fi}
\end{equation}
The cost in the decision-making process is recorded as
\begin{equation}
   G_i(S; {\mathbf{E}}, {\mathbf{c}})= \sum_{i\in{S}} \big\| {\mathbf{c}}\circ ({\mathbf{e}}_j-{\mathbf{e}}_i) \big\|_2,
   \label{Eq:Gi}
\end{equation}
where ${\mathbf{e}}_i$ and ${\mathbf{e}}_j$ represent the endowment vectors of the economies $i$ and $j$, respectively. Based on the definition of the benefit function in Eq.~(\ref{Eq:Fi}) and the cost function in Eq.~(\ref{Eq:Gi}), it can be seen that for two economies to have trade relations, at least one of them needs to be able to profit, that is, the new trade relations must bring positive benefits to one party, as shown below \citep{Yuan-Alabdulkareem-Pentland-2018-NC}:
\begin{equation}
  \Delta u_i (j)= U_i(S_i^{*}; {\mathbf{E}}, {\mathbf{b}}, {\mathbf{c}})-U_i(S_i^{*}/{j}; {\mathbf{E}}, {\mathbf{b}}, {\mathbf{c}}), ~~{\mathrm{if}} ~~ j\in S_i^{*},
  \label{Eq:deltaUi1}
\end{equation}
the above formula represents the incremental utility of economy $i$ when $j$ is among the best trading partners of economy $i$.
\begin{equation}
\Delta u_i (j)= U_i(S_i^{*}\cup {j}; {\mathbf{E}}, {\mathbf{b}}, {\mathbf{c}})-U_i(S_i^{*}; {\mathbf{E}}, {\mathbf{b}}, {\mathbf{c}}), ~~{\mathrm{if}} ~~ j\notin S_i^{*},
\label{Eq:deltaUi2}
\end{equation}
the above formula represents the incremental utility of the economy $i$ when the economy $j$ is not among the best trading partners of the economy $i$.
Substitute Eq.~(\ref{Eq:Fi}) and Eq.~(\ref{Eq:Gi}) into Eq.~(\ref{Eq:deltaUi1}) and Eq.~(\ref{Eq:deltaUi2}) to get:
\begin{equation}
\Delta u_i (j)= \sum_{d=1}^{D_b} b_d\max(e_{jd}-e_{id},0) - \big\| {\mathbf{c}}\circ ({\mathbf{e}}_j-{\mathbf{e}}_i) \big\|_2.
\label{Eq:delta:ui1}
\end{equation}
In the formation of trade cooperation, when the increment $\Delta u_i (j)$ is greater than a certain threshold, the two economies establish a trade relationship.

\subsection{Machine learning model for parameter fitting}

The oil trade decision-making model $U_i(S;{\mathbf{E}},{\mathbf{b}},{\mathbf{c}})$ , which is based on the heterogeneous economy $i$, needs to learn the endowment matrix $\mathbf{E}$ and the weight coefficient ${\mathbf{b}}$ and ${\mathbf{c}}$ from the original iOTN to reconstruct oil trade relationship. Numerical simulations can then be conducted to explore the evolutionary dynamics of the oil trade network, simulate contingency shocks, and evaluate policy effects. Therefore, for the oil trade decision-making model of heterogeneous economies $U_i(S; {\mathbf{E}},{\mathbf{b}},{\mathbf{c}})$, it is vital to obtain the endowment matrix $\mathbf{E}$. A machine learning optimization algorithm is used for parameter learning. Our basic assumption is that a better endowment matrix $\mathbf{E}$ can better reconstruct the original oil trade network based on the oil trade model. The endowment matrix $\mathbf{E}$ and the weight coefficient ${\mathbf{b}}$ and ${\mathbf{c}}$ of the model are learned through the objective optimization function. The objective function of machine learning optimization is the loss function $\mathcal{L}$. A smaller loss function $\mathcal{L}$ indicates a better reconstruction of iOTN based on the endowment matrix $\mathbf{E}$ and the decision model, and also shows that the endowment vector contains the attributes of the economy and the factors that affect trade. The subsequent numerical simulation and policy simulation can be more realistic and close to the real trading environment. The loss function $\mathcal{L}$ takes the following form \citep{Yuan-Alabdulkareem-Pentland-2018-NC}:
\begin{equation}
  \mathcal{L} = \mathcal{L}_{\rm{pos}}  + \mathcal{L}_{\rm{neg}}  + \mathcal{L}_{\rm{fp}}  +\mathcal{L}_{\rm{reg}}.
  \label{Eq:L}
\end{equation}
The smaller the $\mathcal{L}_{\rm{pos}}$ is, the higher the accuracy of predicting the existing connection based on the learned parameters is. A smaller $\mathcal{L}_{\rm{neg}}$ indicates that those non-existent links have a smaller probability of occurring in the prediction process and that the model reconstruction is better. $\mathcal{L}_{\rm{fp}}$ is a penalty item, which is to  penalize the situation where a connection is a non-existent trade relationship, but the model predicts that it is an existing trade relationship. $\mathcal{L}_{\rm{reg}}$ is a regular term. To make the model more stable, when the dimension $D$ of the endowment matrix $\mathbf{E}$ is too large, some unnecessary dimensional values can be made close to zero, which can reduce the dimensionality of the embedding space. The specific details and explanation of $\mathcal{L}_{\rm{pos}}$, $\mathcal{L}_{\rm{neg}} $, $\mathcal{L}_{\rm{fp}}$, $\mathcal{L}_{\rm{reg}}$ can be found in the supplementary materials of the literature \citep{Yuan-Alabdulkareem-Pentland-2018-NC}. In order to find the minimum loss function $\mathcal{L}$, the optimization model of machine learning is as follows
\begin{equation}
\begin{aligned}
\hat{\mathbf{E}},\hat{\mathbf{b}},\hat{\mathbf{c}}&=\arg\min_{\mathbf{E},{\mathbf{b}}, {\mathbf{c}}}\mathcal{L}({\mathbf{E}},{\mathbf{b}}, {\mathbf{c}}|{\mathbf{A}})\\
    {\mathrm{subject~to}}:~~~  c_d\geq 0,\forall d&=1,2,...,D\\
                               \sum_{i=1}^{N}e_{id}&=0,\forall d=1,2,...,D\\
                               \big\| {\mathbf{E}}_{:d} \big\| _2^2&=N,\forall d=1,2,...,D
\label{Eq:Minimize}
\end{aligned}
\end{equation}
In the model training process, the selection of hyperparameters refers to the parameter settings in the literature \citep{Yuan-Alabdulkareem-Pentland-2018-NC}, the optimization process is iterated 10,000 times, and the learning rate is 0.01. We averaged the endowments among economies. The endowments of all economies in the same dimension have a mean value of zero and are standardized. Therefore, the endowment vectors of different economies in learned $\mathbf{E}$ are comparable.

\section{Calibration results}
\label{S1:EmpAnal}

\subsection{The spatial distribution of economies in endowments space}

\begin{figure*}[!t]
\centering
\includegraphics[width=0.96\linewidth]{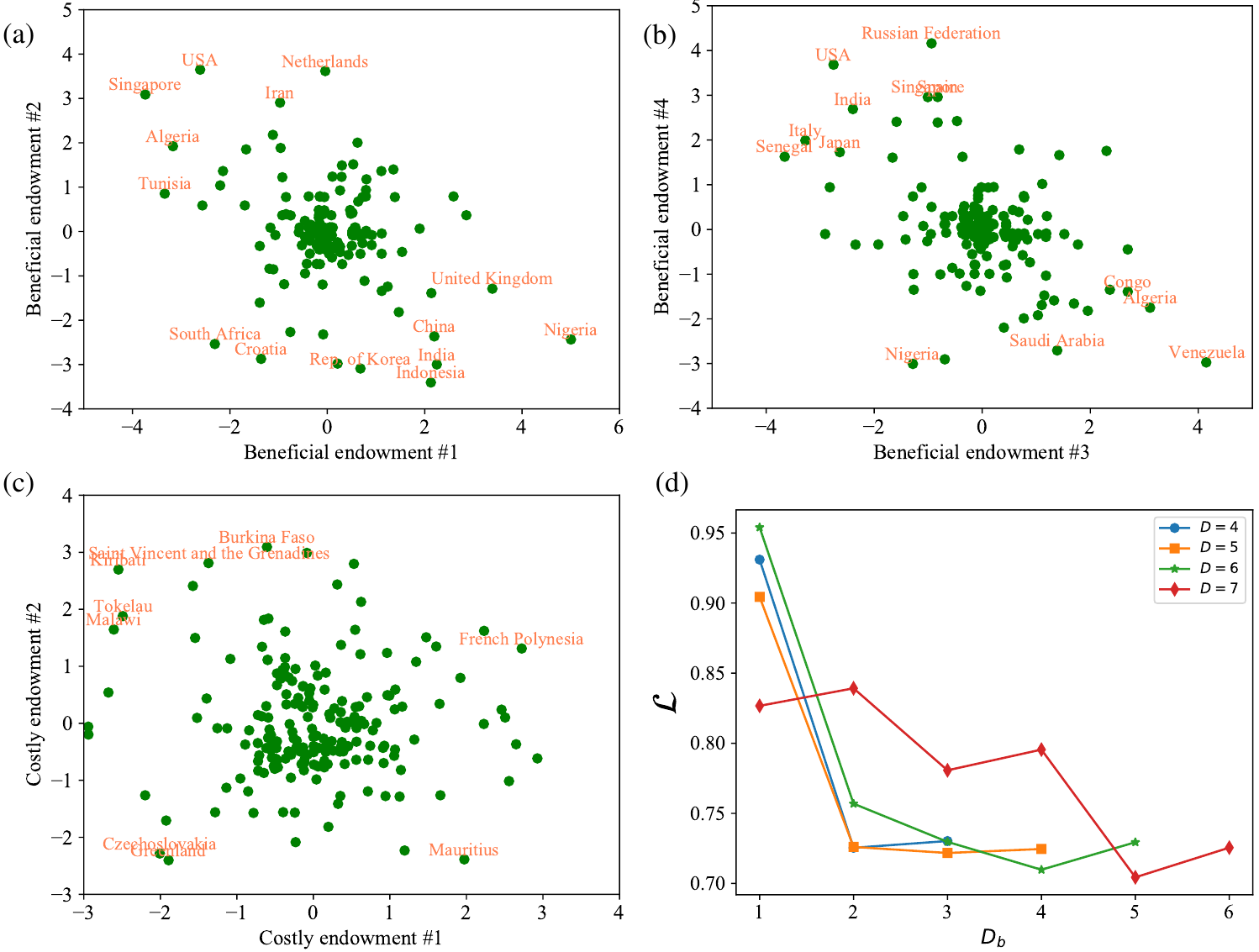}
  \caption{
The endowments in the case of the embedded dimension $D=6$ and the benefit endowment dimension $D_b=4$. a) The first and second columns of the benefit endowment of economies in the iOTN. b) The third and fourth columns of benefit attributes. c) The first and second columns of the cost attributes. d) The relationship between the value of the objective function $\mathcal{L}$ and the benefit endowment dimension $D_b$ under different embedded dimensions $D$.
  }
    \label{Fig:oil:endowment}
\end{figure*}

The analysis of oil trade decision-making process based on the endowment vectors of heterogeneous economies has some interpretability. We find that economies have different trade endowments. To be able to visualize and interpret the learned endowment feature vectors, we draw the endowment vectors of the economies as scatter plots for visualization, as shown in Fig.~\ref{Fig:oil:endowment}(a,b,c).

Fig.~\ref{Fig:oil:endowment}(a,b,c) show the endowment vector of the economy with the embedding dimension $D=6$ and the benefit endowment dimension $D_b=4$. The major energy demand and supply economies such as the USA, the United Arab Emirates, Nigeria, Indonesia, China, and Singapore are in a relatively remote location. Their endowments are different from other economies. We have to admit that it is difficult to explain each endowment dimension in an economic sense. Because the coordinate vector embedded as a variable contains many unmeasured trade factors, so it is basically impossible for a certain indicator to be completely correlated with the economic variables in reality.

The embedded dimension is directly related to the decision model's decision variables, including the benefit and cost endowment variables. In order to select the most suitable endowment dimension, we have chosen different embedding dimensions $D=4,5,6,7$ and the corresponding benefit endowment dimension $D_b$. In order to obtain a smaller loss function value during the optimization process, we calculated the loss value $\mathcal{L}$ for different benefit dimensions $D_b$. Fig.~\ref{Fig:oil:endowment}(d) shows the objective function value $\mathcal{L}$ under different embedding dimension $D$ and benefit endowment dimension $D_b$. It can be seen that the higher the dimension is, the smaller the loss function is. In Fig.~\ref{Fig:oil:endowment}(d), each solid line represents a case of the dimension $D$. The abscissa represents the benefit dimension with the size $D_b$, and the ordinate represents the loss function value $\mathcal{L}$. The lowest point of each solid line can be used to determine the optimal benefit dimension $D_b$. For the iOTN, when the embedded dimensions are $D=4,5,6,7$, the optimal benefit dimensions are $D_b^{*}=2,3,4,5$ respectively. In the subsequent analysis, under the corresponding embedding dimension, we only considered the case with the optimal benefit dimension $D_b^{*}$.

The coordinates in Figure 2(a,b) indicate the benefit endowments, while Figure 2(c) indicates the cost attributes. The economies marked in Figure 2 are those with very large attribute values. The larger the cost attribute of the economies in (c), the more difficult it is to establish trade relations with other economies. While the larger economies in Figure 2(a,b) correspond to some economies with larger trade volumes. Thus, the inconsistency of the economies shown in Figure 2 is another indication that the algorithm identifies the heterogeneity among the economies. Figure 2 intuitively shows the interpretability of the feature vectors of the economies. However, interpretability has limitations. The feature vectors cannot fully correspond to some observable economic variables or attribute variables of the economies.

\subsection{Interpretability analysis of endowments}

The black box problem of machine learning and deep learning, namely interpretability, has always been challenging, and it is impossible to decompose the endowment vector's connotation by a simple method. Therefore, the limitations of the method used in the paper are also reflected here. In order to further explore the specific economic meaning of endowment vectors, correlation analysis can be combined with existing economic variables. We attempt to define some measurements based on the endowment vector to discover some characteristics in the iOTN and provide more options for analyzing the network.

We define two measurements based on the endowment vector to fathom more meaningful economical information from the endowment vector. First, we calculate the economy’s oil trade power index \citep{Yuan-Alabdulkareem-Pentland-2018-NC}:
$
{\rm{Power}}(i)=\sum_{d=1}^{D_b}b_d e_{id}.
$

In order to characterize the exclusion of different economies, an indicator of economy exclusion is defined, which is the characteristic that other countries do not tend to trade with an economy or the economy itself does not trade with other economies \citep{Yuan-Alabdulkareem-Pentland-2018-NC}. The economic exclusion index is defined as
$
{\rm{Exclusion}}(i)=\left[\sum_{d=D_b+1}^{D}(c_d e_{id})^2\right]^{\frac{1}{2}}.
$
The exclusion or power is informative for other economies. We can study the power and exclusion of economies to provide recommendations for the economies' oil trade decisions. In order to verify the validity of trade exclusion and trade power measurement, we analyze the Spearman correlation coefficient between the trade network structure variables and exclusion and power \citep{Best-Roberts-1975-AppliedStatistics}.

\textbf{Degree centrality} measures the number of oil trading partners of an economy $i$ and is defined as
$
{\rm{Degree}}(i)=\sum_{j=1}^{N}a_{ij}.
$

\textbf{Closeness centrality} is used to measure the average distance of an economy to other economies and defined as
$
{\rm{Closeness}}(i)=\frac{N-1}{C_i}.
$
$N$ is the number of economies in the network, and $C_i$ is the sum of the distances from economy $i$ to $N-1$ other economies. For isolated economy $i$, ${\rm{Closeness}}(i)$ is 0.

\textbf{Clustering} is a measurement of local centrality. The clustering coefficient measures a density of triangular trade relations in trade relations between economies. The clustering coefficient is defined as
$
{\rm{Clustering}}(i) = \frac{\sum_{m,j=1}^{N}a_{im}a_{mj}a_{ji}}{{\rm{Degree}}(i)({\rm{Degree}}(i)-1)},
$
where $a_{ij} = 1$ and $a_{ij} = 0$ correspond to connection
and disconnection for nodes $i$ and $j$. 

\begin{table*}[!t]
  \centering
  \caption{
The correlation between exclusion, power and network degree, closeness centrality, clustering centrality, export volume, and import volume. The bold correlation coefficient is not significant when the significance level is 0.05. From top to bottom, the embedding dimensions are $D=4,5,6,7$, and the optimal benefit endowment dimensions are set as the corresponding $D_b^{*}$.}
\begin{tabular}{lrrrrrrr}
\toprule
{} &  Power &  Exclusion &  Import &  Export &  Degree &  Closeness &  Clustering \\
\midrule
Panel A: $D=4$ \\
Power     &   1.00 &      -0.35 &    0.44 &    0.38 &    0.42 &       0.40 &       -0.58 \\
Exclusion &  -0.35 &       1.00 &   -0.68 &   -0.82 &   -0.90 &      -0.89 &       \textbf{-0.04} \\
\midrule
Panel B: $D=5$ \\
Power     &   1.00 &      -0.33 &    0.49 &    0.46 &    0.50 &       0.48 &       -0.55 \\
Exclusion &  -0.33 &       1.00 &   -0.67 &   -0.81 &   -0.87 &      -0.87 &       \textbf{-0.06} \\

\midrule
Panel C: $D=6$ \\
Power     &   1.00 &      -0.29 &    0.56 &    0.53 &    0.58 &       0.55 &       -0.55 \\
Exclusion &  -0.29 &       1.00 &   -0.62 &   -0.78 &   -0.84 &      -0.85 &       \textbf{-0.12} \\

\midrule
Panel D: $D=7$ \\
Power     &   1.00 &      -0.25 &    0.60 &    0.59 &    0.66 &       0.63 &       -0.47 \\
Exclusion &  -0.25 &       1.00 &   -0.57 &   -0.72 &   -0.78 &      -0.79 &       -0.20 \\
\bottomrule
\end{tabular}
\label{Tab:oil:corr}
\end{table*}

The correlations between power and exclusion, oil import, export and network structure measurements are shown in Table~\ref{Tab:oil:corr}. First, trade power is negatively correlated with trade exclusion, and the absolute value of the correlation coefficient decreases with the increase of $D$. Trade power is positively correlated with import and export volume, the number of trading partners, closeness centrality, and clustering coefficient, and the correlation coefficient increases with an increase of $D$. Trade exclusion is negatively correlated with import and export volume, the number of trading partners, and closeness centrality, and the absolute value of the correlation coefficient decreases as $D$ increases. Trade exclusion and clustering coefficient show a weak negative correlation. The absolute value of the correlation coefficient increases as $D$ increases.

According to the results from different embedding dimensions, it can be seen that the correlation between the structural variables of the original network and economic power and exclusion are robust. The results also prove that the endowment vector learned by the optimization algorithm has  economic implications, which can be used to measure the attributes of the economy and guarantees the 
validity of subsequent policy simulations.

\subsection{Model validation}
\label{S1:ModelCalibration}

\begin{figure*}[!t]
   \centering
   \includegraphics[width=0.999\linewidth]{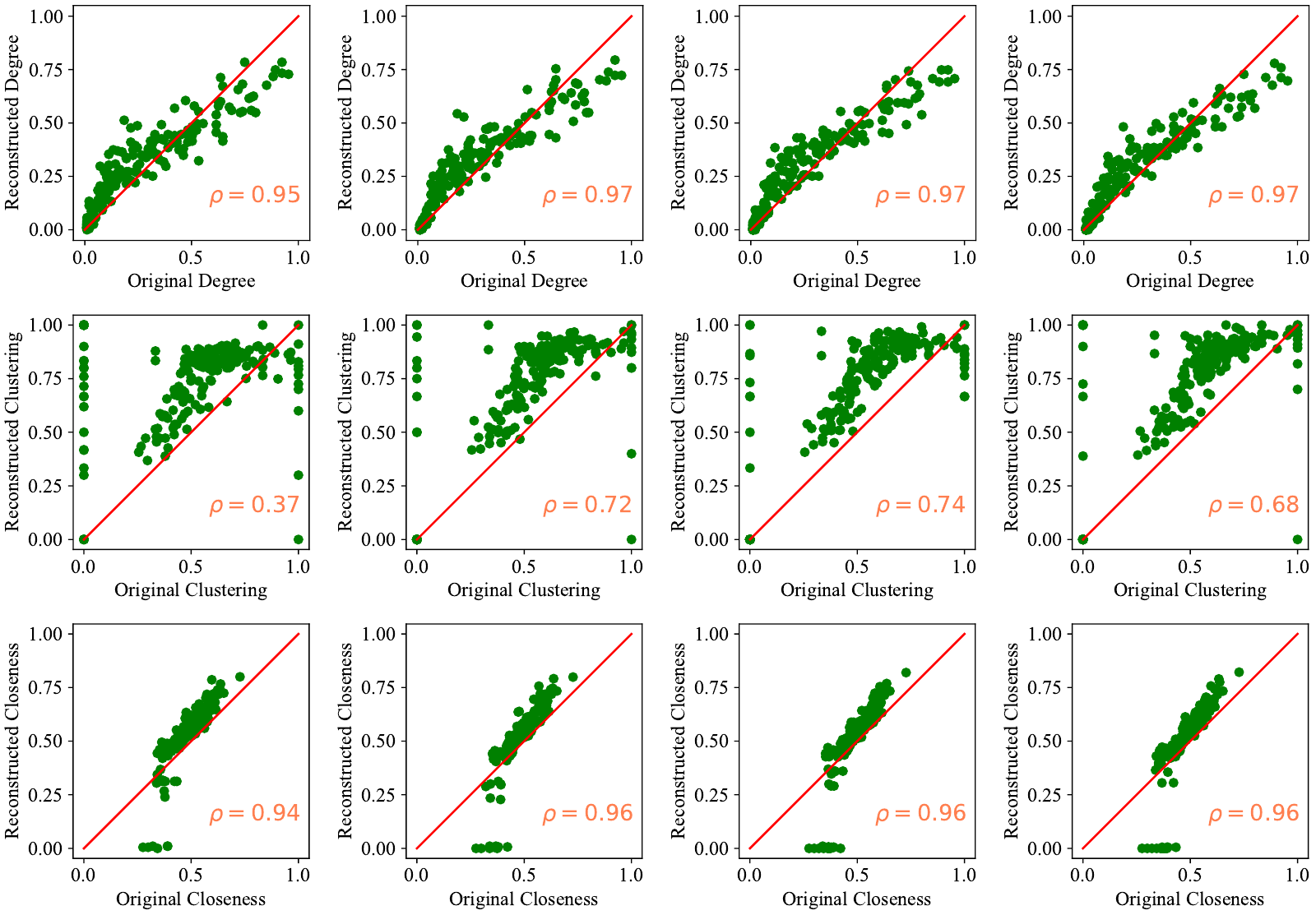}
   \caption{
In the embedding spaces with different dimensions, the scatter plots of economies' network degree, clustering coefficient, and closeness of the restructured network and the original network. The rank correlations between the reconstructed network's structural variables and the original network are shown in the plots. There is a comparison for the network degree, clustering coefficient, and closeness from top to bottom. From left to right, the embedding dimensions are $D=4, 5, 6, 7$. 
   }
   \label{Fig:oil:validation}
\end{figure*}

To further verify that the endowment vectors of economies have certain practical significance, we reconstructed the network based on the learned model parameters to verify whether some structural measurements in the reconstructed network have some similarity with the original network.

We use Eq.~(\ref{Eq:delta:ui1}) to calculate $\Delta_{ij}={\rm{min}}(\Delta u_i (j),\Delta u_j (i))$, and take the $N_{\rm{edges}}$ edge with the largest value of $\Delta_{ij}$ to reconstruct the $N_{\rm{edges}}$ trade relations in the original iOTN. $N_{\rm{edges}}$ is the number of original trade relationships in iOTN. Then, we statistically analyze the network structure parameters such as the network degree of the economy, the clustering coefficient, and the closeness. They are respectively denoted as reconstructed ${\rm{Degree}}(i)$, reconstructed ${\rm{Clustering}}(i)$, reconstructed ${\rm{Closeness}}(i)$.

To quantify the performance of reconstruction, Spearman's correlation coefficient is used to measure the rank correlation between the structural variables of the reconstructed iOTN and that of the original iOTN \citep{Best-Roberts-1975-AppliedStatistics}.
\begin{equation}
    \rho = 1 - \frac{6\sum_{i=1}^{N}\big(X_i-Y_i\big)^2}{n(n^2-1)},
\end{equation}
where $X_i$ and $Y_i$ are the ranks of the two variables sorted by size.

In Fig.~\ref{Fig:oil:validation}, the network degree, clustering coefficient, and closeness of the reconstructed network are highly correlated with the original network, which shows that the endowment vector learned by qualitative individuals can reflect the hidden variables of the economy in the iOTN and provides a guarantee for the effectiveness of our subsequent simulation. From left to right in Fig.~\ref{Fig:oil:validation}, the embedding dimensions are $D=4,5,6,7$, and the optimal benefit endowment dimensions are their respective $D_b^{*}$. The comparison shows that the higher the dimension is, the better the correlation is, and the better the structure and properties of the original network can be reproduced. 

\section{Policy simulation of oil trade}
\label{S1:PolicySimulation}

It can be concluded from the interpretability analysis of the endowment attributes and the verification of the model that  the model has learned the characteristics of the iOTN. The adaptive decision-making model of an economy has a certain validity. Based on the endowment of economies, it is feasible to conduct policy simulations on decision-making models. Changing the model parameters can simulate different market environments' changes, the evolution of the iOTN, and the future development of global oil trade.

With the trade endowment of the economy learned, the adaptive decision model of the economy is able to simulate the evolution and development patterns of trade networks under different trade environments. For example, the cost importance parameter ${\mathbf{c}}$ in the model can be adjusted to simulate the oil trade situation after the occurrence of events such as the COVID-19 epidemic, trade wars, trade barriers, and trade tariffs, which all lead to increased trade costs. The new cost importance parameter is defined as
\begin{equation}
{\mathbf{c}}_{\alpha}=(1+\alpha){\mathbf{c}}.
\label{Eq:c:new}
\end{equation}

We use the parameter $\alpha$ to measure the change of the model parameter ${\mathbf{c}}$. When ${\mathbf{c}}_{\alpha}$ increases ($\alpha>0$), it corresponds to the increase in trade costs. When ${\mathbf{c}}_{\alpha}$ decreases ($\alpha<0$), it corresponds to the reduction of trade costs, such as reducing tariffs and zero tariffs. The value of $\alpha$ is between -0.4 and 0.4, and the iOTN is re-simulated based on the following formula:
\begin{equation}
\Delta u_i (j)= \sum_{d=1}^{D_b} b_d\max(e_{jd}-e_{id},0) - \big\|{\mathbf{c}}_{\alpha}\circ ({\mathbf{e}}_j-{\mathbf{e}}_i) \big\|_2.
\label{Eq:delta:ui1:new}
\end{equation}
When $\alpha=0$, that is, ${\mathbf{c}}_{\alpha}=\mathbf{c}$, we calculate $\Delta_{ij}={\rm{min}}(\Delta u_i (j),\Delta u_j (i))$, and take the first $N_{\rm{edges}}$ edge with the largest value in $\Delta_{ij}$ to reconstruct the trade relations. $N_{\rm{edges}}$ is the number of original trade relationships in iONT. We will use these $N_{\rm{edges}}$ edges to find the minimum value of $\Delta_{ij}$ and record it as $\Delta_{\rm{min}}$. $\Delta_{\rm{min}}$ can be regarded as the utility threshold value for new oil trade relations in the iOTN. Then, we statistically analyze the network structure parameters, such as the network degree of the economy, the clustering coefficient, and the closeness. They are respectively denoted as reconstructed ${\rm{Degree}}(i)$, ${\rm{Clustering}}(i)$, ${\rm{Closeness}}(i)$.

When $\alpha=-0.4,...,0.4$, we recalculate $\Delta_{ij}$. When the trade relationship corresponds to $\Delta_{ij}$ satisfy
\begin{equation}
\Delta_{ij}={\rm{min}}(\Delta u_i (j),\Delta u_j (i))>\Delta_{\rm{min}}.
\label{Eq:new:delta:c}
\end{equation}
The corresponding economies $i$ and $j$ establish oil trade relations. Statistical analysis of the structural variables of the reconstructed network, such as the network degree of the economy, the clustering coefficient, and the closeness are recorded as ${\rm{Degree}}^{\alpha}(i)$, ${\rm{Clustering} }^{\alpha}(i)$,
${\rm{Closeness}}^{\alpha}(i)$.

Adjusting the cost importance parameter $\mathbf{c}$ has a different effect on the structural measurements of trade network, such as degree, clustering coefficient, and closeness. The changes in degree affect the number of trade partners of the economy. Under trade frictions or trade shocks, the changes in the number of partners can be used to measure the vulnerability of the economy. When the amount of change is large, the vulnerability is relatively large. The vulnerability based on degree, clustering, and closeness is defined as
\citep{Dall'Asta-Barrat-Barthelemy-Vespignani-2006-JSM,Yuan-Alabdulkareem-Pentland-2018-NC} 
\begin{equation}
V_{\rm{Degree}}=\frac{{\rm{Degree}}^{\alpha}(i)}{{\rm{Degree}}(i)},~~~~~~V_{\rm{Clustering}}=\frac{{\rm{Clustering}}^{\alpha}(i)}{{\rm{Clustering}}(i)},~~~~~~V_{\rm{Closeness}}=\frac{{\rm{Closeness}}^{\alpha}(i)}{{\rm{Closeness}}(i)},
\label{Eq:Vulnerability}
\end{equation}
where ${\rm{Degree}}(i)$, ${\rm{Clustering}}(i)$, ${\rm{Closeness}}(i)$ respectively represent the restructured network's degree, clustering coefficient, and closeness when $\alpha=0$.

\begin{figure}[!t]
\centering
\includegraphics[width=0.98\linewidth]{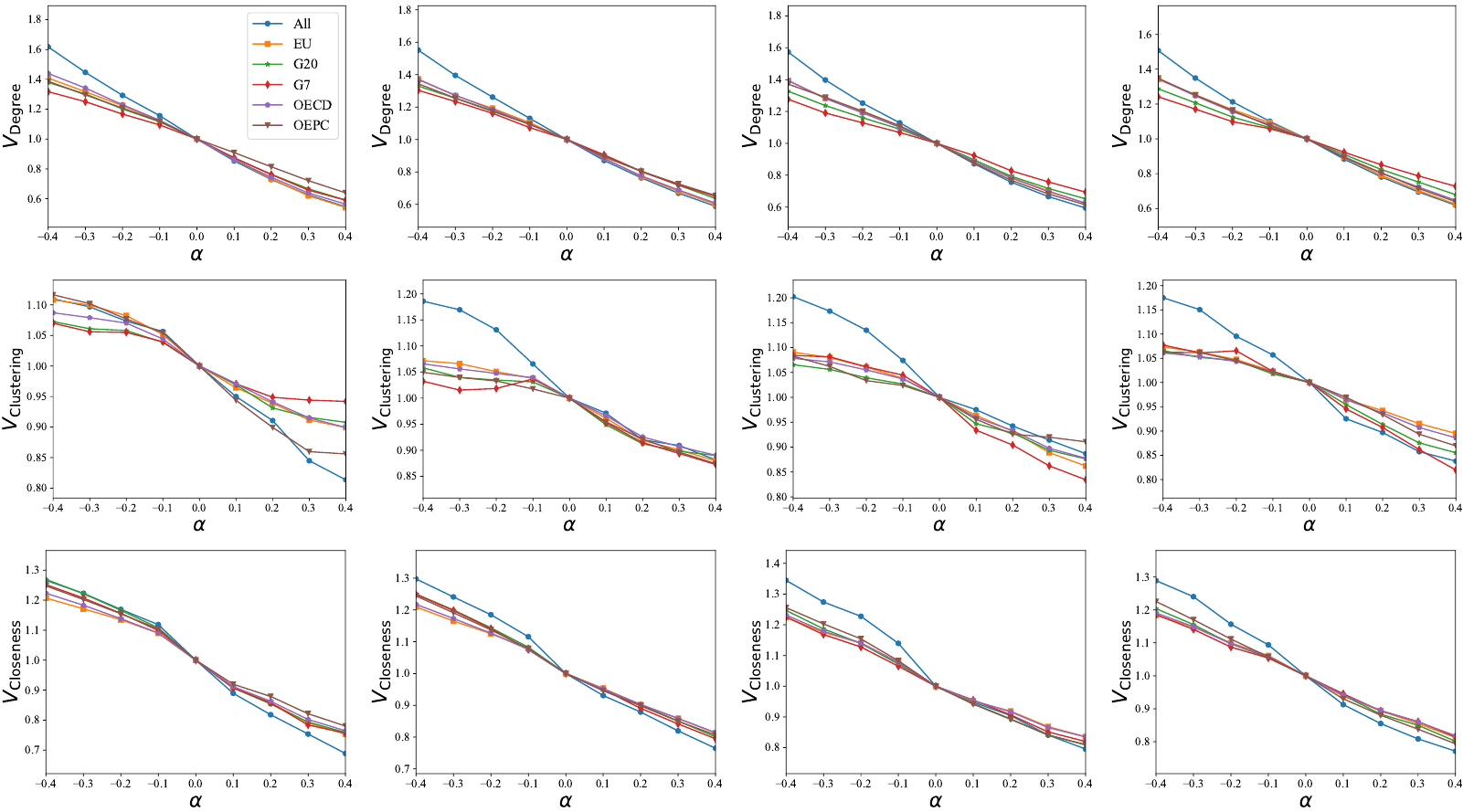}
  \caption{
 The vulnerability $V$ of the economy under the influence of trade friction factors $\alpha$. Under different $\alpha$ values, there are the average network degree, the average clustering coefficient, and the average closeness from top to bottom. From left to right, the embedding dimensions are $D=4,5,6,7$, and the optimal benefit endowment dimensions are their respective $D_b^{*}$.
  }
    \label{Fig:oil:abm}
\end{figure}

Different cost coefficients $\alpha$ have different effects on network structure variables. Fig.~\ref{Fig:oil:abm} shows the impact of cost coefficient changes on the three network structure variables of the international organizations in the iOTN, including the degree, clustering coefficient, and closeness. The network degree can be used to measure the prosperity of the global energy trade network. A larger network degree indicates greater edge density, more trade relations, and more trade activities. The clustering coefficient is used to measure the density of triangular trade relations between economies. A larger clustering coefficient indicates a larger number of triangular oil trade relations in the iOTN. The closeness characterizes the degree of connection between trade network economies. If the closeness is greater, the path between the two economies will be shorter, and this is more conducive to the effectiveness of oil trade.

In Fig.~\ref{Fig:oil:abm}, the change of cost coefficient $\mathbf{c}$ has a greater impact on the average degree of international organizations. When $\mathbf{c}_{\alpha}=0.6\mathbf{c}$, the average degree of the entire network has increased by more than 60\%. The increase and decrease of the trade cost coefficient $\mathbf{c}_{\alpha}$ have an asymmetric effect on the average degree of the network. When ${\mathbf{c}}_{\alpha}=1.4{\mathbf{c}}$, the average degree of the entire network is reduced by nearly 40\%. The reduction of $\mathbf{c}_{\alpha}$ ($\alpha<0$) corresponds to the lifting of trade wars, the lifting of trade barriers, and the reduction of trade tariffs. The impact of such policies varies across organizations, with the average degree of impact being highest for all global economies, followed by the OECD, and the G7 having the least impact. In general, the impact of trade cost on powerful economies is less than that on weak economies, and the impact on OPEC is greater than that on G7. The increase of $\mathbf{c}_{\alpha}$ ($\alpha>0$) corresponds to situations such as trade wars, trade barriers, and trade tariffs. Such policies that increase trade costs have less impact on the economy compared to policies that reduce trade costs. The small difference in the impact on different organizations suggests a slower recovery in trade.

\begin{figure*}[!t]
\centering
\includegraphics[width=0.999\linewidth]{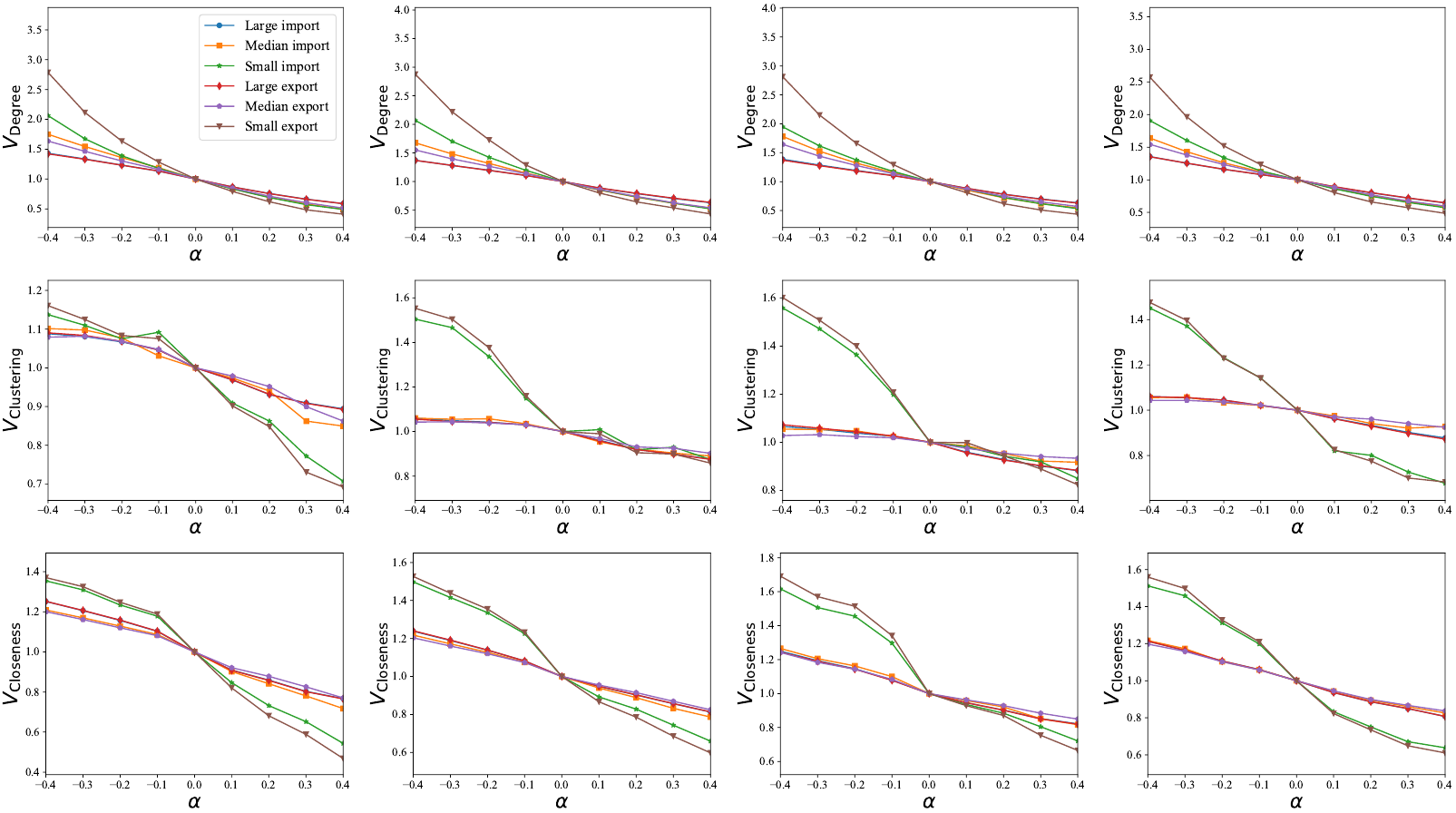}
  \caption{
The vulnerability analysis of economies in large, medium, and small groups under the influence of trade friction factors $\alpha$. According to the volumes of oil export and import, the economies can be divided into large, medium, and small ones. The settings in the figure are the same as those in Fig.~\ref{Fig:oil:abm}.
  }
    \label{Fig:oil:abm:groups}
\end{figure*}

In Fig.~\ref{Fig:oil:abm}, the change of the cost coefficient $\mathbf{c}$ has a smaller effect on the average clustering coefficient of the economies in the organization than on the average degree. When $\mathbf{c}_{\alpha}=0.6\mathbf{c}$, the average clustering coefficient of the entire network increases by no more than 15\%. The increase and decrease of the trade cost coefficient $\mathbf{c}_{\alpha}$ also have an asymmetric effect on the average clustering coefficient. When $\mathbf{c}_{\alpha}=1.4\mathbf{c}$, the average clustering coefficient of the entire network is reduced by approximately 15\%. Similar to the previous results, changes in trade costs have less impact on powerful economies than on weak ones. The closeness centrality of different organizations are affected similarly, and strong economies are also less affected than weak ones. These results all point to the increased risks and greater vulnerability of smaller economies in the crisis and the additional challenges they will encounter in the later recovery process. Since $\alpha$ is greater than 0, it means that trade frictions and trade costs are increased, so trade relations will be reduced, so vulnerability $V$ is less than 1. However, all curve contains all economies, and it contains a large number of small economies, which are more affected by trade shocks, so the curve is smaller than that of other organizations, which contain a large number of powerful economies. Stronger economies are also less affected than weaker ones, resulting in a $V$ value close to 1.

To verify whether smaller economies are more affected in the simulation, we divided the economies into three groups according to the oil trade volume of export and import. For statistical validity, we divided all economies into three groups, each containing the same number of economies. Fig.~\ref{Fig:oil:abm:groups} shows the changes in the average network degree, average clustering coefficient, and average closeness of economies in different groups under different $\alpha$ values. From left to right in Fig.~\ref{Fig:oil:abm:groups}, the embedding dimensions are $D=4,5,6,7$, and the optimal benefit endowment dimensions are their respective $D_b^{*} $. There are the average network degree, the average clustering coefficient, and the average closeness from top to bottom. As shown in Fig.~\ref{Fig:oil:abm:groups}, the traditional powerful oil import and export economies will be less affected when they encounter shocks. The economies with small oil imports and exports are more susceptible to shocks. Further comparison and analysis of Fig.~\ref{Fig:oil:abm:groups} shows that oil-exporting economies are more susceptible to shocks. The influence of the importing economies is relatively small. To a certain extent, this finding also shows that different economies need to diversify their strategies to respond to shocks in the face of changes in the international environment. Risk prevention and risk mitigation should be based on their trade attributes. Similarly, we can use simulation to study individual economies' impact when external emergencies or disasters occur. The simulation can provide certain predictions and preliminary preparations for disaster crises. In Fig.~\ref{Fig:oil:abm:groups} we can also find that under different trade cost coefficients $\mathbf{c}_{\alpha}$ value, different embedded dimension simulation, and different structural variables, the vulnerability has a relative similarity. This finding illustrates to a certain extent that the numerical simulation results of our method have certain robustness. It further illustrates that our method has certain credibility for measuring the endowments of different economies.

\section{Discussion and application}
\label{S1:Conclude}

The importance and vulnerability of the oil trade necessitates a move toward diversification and efficiency. We constructed an undirected weighted international oil trade network based on the oil trade data from 1990 to 2019. The main contribution of this paper is that the iOTN is reconstructed  by integrating game theory and utility theory, the trade network data being directly put into the machine learning algorithm as a data representation. Based on the network representation learning, the large-scale optimization method of machine learning is used to perform directly representation learning on heterogeneous economies in the iOTN. Simulation based on game theory and agent-based model provides suggestions for selecting trade partners and the formulation of trade cooperation for various economies. The main finding is that export-oriented economies have a greater impact than import-oriented economies after receiving external shocks in the iOTN. The impact of external events that leads to an increase or decrease in the cost of trade friction has an asymmetric impact on the iOTN, and the reduction in trade friction has a greater impact on oil trade. Smaller economies are more severely affected when they are hit by external events than larger economies.

For international organizations, the dynamic evolution trend of international oil trade relations can be obtained through numerical simulation. For individual economies, it is possible to find cooperation strategies in their own economies' international trade status under specific trading environment changes. From the numerical simulation results, it can be found that different economies have to formulate and respond to different trade strategies based on their characteristics and cannot merely copy the trade response of other economies. Under our model framework, more detailed simulation and analysis can be conducted for individual economies. For example, the trade cost coefficient of a particular economy can be changed separately, instead of changing the overall cost coefficient as in this article. In addition, when an economy becomes politically unstable, the increase in its cost coefficient will definitely impact other economies. Our finding shows that there is a need for more cooperative relationships between different economies to diversify their trading partners to increase their ability to resist and mitigate risks. Based on the results of our vulnerability analysis, we can give early warning signals to some of the more vulnerable economies, and those economies can make trade strategy adjustments in advance. Considering the learned attributes of economies, we can classify economies into different groups, so that we can determine trade object attributes more precisely and specify trade strategies to provide trade efficiency and stability.

Follow-up work can integrate more economic entity endowment data, conduct early warning of systematic risk and critical risk path identification, and propose more accurate response strategies. Based on the heterogeneous subject's adaptive decision model, the pattern of global energy trade operation can be monitored dynamically and in real time. Integrating emergencies (national bankruptcy, economic sanctions, and epidemic-affected economies), computational experiment methods and ``scenario-response'' models can make the simulated ``scenarios'' more in line with the real domestic trade system settings. Risk warning and risk response strategy evaluation will be conducted, corresponding plans will be made, and scientific and feasible strategies will be formulated.







\section*{Declaration of competing interest}

The authors declare that they have no conflict of interest.

\section*{Data Availability}

Oil data sets related to this article can be found at https://comtrade.un.org/, an open-source online data repository. The geographic distance data sets of the CEPII database come from http://www.cepii.fr/.


\end{document}